\documentclass[12pt, a4paper]{article}\usepackage[]{graphicx}\usepackage[]{xcolor}
\makeatletter
\def\maxwidth{ %
  \ifdim\Gin@nat@width>\linewidth
    \linewidth
  \else
    \Gin@nat@width
  \fi
}
\makeatother

\definecolor{fgcolor}{rgb}{0.345, 0.345, 0.345}

\usepackage{framed}
\makeatletter
 {\par\unskip\endMakeFramed%
 \at@end@of@kframe}
\makeatother

\definecolor{shadecolor}{rgb}{.97, .97, .97}
\definecolor{messagecolor}{rgb}{0, 0, 0}
\definecolor{warningcolor}{rgb}{1, 0, 1}
\definecolor{errorcolor}{rgb}{1, 0, 0}
\newenvironment{knitrout}{}{} % an empty environment to be redefined in TeX

\usepackage{alltt}
\usepackage{verbatim}
\usepackage[vmargin=3cm, hmargin=2cm, headheight=14pt]{geometry}
\usepackage{url}
\usepackage{hyperref}
\usepackage{fancyhdr}
\usepackage{color}
\usepackage{xcolor}
\usepackage{amsmath, amssymb}
\usepackage{longtable}
\usepackage{lscape}
\usepackage[numbers]{natbib}
\usepackage{xspace}
\usepackage{booktabs}
\usepackage[font=sf, labelfont={sf}, margin=1cm]{caption}
\usepackage[T1]{fontenc}
\usepackage[utf8]{inputenc}
\usepackage{textcomp}
\usepackage{authblk}

\usepackage{times} % Times Font

\newcommand{\pt}{{p_T}} % Probability of being in treatment group
\newcommand{\pc}{{p_C}} % Probability of being in control group
\newcommand{\ptie}{{p_{Tie}}} % Probability of a tie
\newcommand{\WR}{{\text{WR}}} % Win Ratio
\newcommand{\niter}{{n_{sim}}} % Number of iterations
\newcommand{\semc}{{\se_{\scriptscriptstyle{\text{MC}}}}} % MC standard error

\DeclareMathOperator{\se}{se}   % standard error
\DeclareMathOperator{\Var}{Var} % variance
\DeclareMathOperator{\Wb}{Wb} % Weibull distribution

\IfFileExists{upquote.sty}{\usepackage{upquote}}{}
\begin{document}

% ========================== Title and Authors =================================
\title{The Win Ratio at the Design Stage of Clinical Trials}

\author[1]{David Kronthaler}
\author[2]{Matthias Schwenkglenks}
\author[3]{Felix Beuschlein}
\author[1]{Ulrike Held}

\affil[1]{Epidemiology, Biostatistics and Prevention Institute, Department of Biostatistics, University of Zurich, Zurich, Switzerland}

\affil[2]{Health Economics Facility and Institute of Pharmaceutical Medicine, Department of Public Health, University of Basel, Basel, Switzerland}

\affil[3]{Department of Endocrinology, Diabetology and Clinical Nutrition, University Hospital Zurich and University of Zurich, Switzerland}

% \corres{*Ulrike Held, \email{ulrike.held@uzh.ch}}
% 
% \presentaddress{Hirschengraben 84, 8001 Zurich, Switzerland}

\maketitle
\newpage
% ============================== Abstract ======================================
\subsection*{Abstract}
The win ratio offers a flexible approach to incorporate the hierarchy of clinical outcomes into the analysis of a composite endpoint, enabling simultaneous consideration of multiple outcome types, unlike traditional time-to-first-event (TTFE) analysis or focus on a single outcome. We examined the statistical power of the win ratio compared to single-endpoint analyses and TTFE analysis through a case study and simulation studies. Furthermore, we provide a novel formula to estimate the required sample size for win ratio analysis based on the desired width of its confidence interval, facilitating precision-based trial design.

Our results indicate that win ratio analysis generally outperforms single-endpoint analyses when treatment effects on lower-ranked outcomes are moderate compared to those on higher-ranked outcomes. The win ratio can provide greater power than TTFE analysis, especially when the effect on the highest-ranked outcome is substantial, reaching increases in power up to 50\%. Further, even for moderate treatment effects on the highest-ranked outcome, win ratio analysis achieved higher power.

Future work should expand our simulations to additional data-generating mechanisms and outcome types, particularly ordinal outcomes, where the win ratio provides an alternative to existing non-parametric and parametric methods.

Our findings highlight the potential of the win ratio to improve statistical efficiency in pharmaceutical and other clinical trial designs using composite endpoints, particularly when no single component dominates the treatment effect. However, when continuous outcomes occupy the top of the hierarchy, these tend to drive overall analysis, sidelining contributions of lower-ranked outcomes and limiting benefits of hierarchical win ratio analysis.

\subsection*{Keywords}
win ratio, composite endpoint, power, precision, trial design, simulation
\newpage

% ============================== Main text =====================================
\section{Introduction}\label{sec:intro}
Traditional analyses of composite endpoints in many drug development trials or other clinical areas typically focus on the time to the first occurrence of any of several clinically related events \cite{redfors2020win, monzo2024use}. These analyses usually employ standard survival analysis methods such as Cox proportional hazards (PH) regression \cite{cox1972regression} or Kaplan-Meier \cite{kaplan1958nonparametric} or Turnbull \cite{turnbull1974nonparametric} curves to illustrate differences in time-to-first-event (TTFE) distributions between treatment groups \cite{redfors2020win}. 

However, TTFE analysis does not account for the clinical hierarchy of outcomes based on severity and importance for patience \cite{redfors2020win}. For example, in a composite endpoint including time to death, hospitalization, and any adverse event, TTFE considers only the first event to occur --- regardless of its clinical relevance. TTFE analysis also ignores recurrent events, such as repeated hospitalizations, which can significantly impact patient outcomes and healthcare resources. Moreover, events that occur earlier, like hospitalizations, are often less severe than those occurring later, such as death. As a result, TTFE analysis may underrepresent more severe outcomes, potentially resulting in a misleading interpretation of treatment effects.

Competing risk analysis offers an alternative framework that addresses some limitations of standard time-to-event (TTE) methods by recognizing that certain events may preclude others \cite{satagopan2004note}. A competing risk is an event that either hinders or alters the probability of another event occurring --- for example, death may prevent subsequent hospitalization. Traditional TTE methods can yield biased estimates in this context, as they typically treat competing events as non-informative censoring. More appropriate approaches include the cumulative incidence function and Fine and Gray’s subdistribution hazard model \cite{fine1999proportional}. However, competing risk analysis, like TTFE, treats all events in the composite as equally important, ignoring their clinical hierarchy. In addition, both competing risk and TTFE analyses are limited to TTE outcomes, making it difficult to incorporate outcomes of other types (e.g., continuous biomarkers or quality-of-life measures) into a single composite endpoint.

Recognizing these limitations, attention has been directed towards methods that account for the hierarchical structure of composite endpoints while accommodating outcomes of different types \cite{Wang2016, li2025win}. Approaches such as the Finkelstein--Schoenfeld method \cite{finkelstein1999combining}, Buyse's generalized pairwise comparisons \cite{buyse2010generalized} and the win ratio (WR) \cite{Pocock2011} have been proposed to address this issue. Among these, WR analysis has gained the most traction and is increasingly used for (re-)analysis of pharmaceutical and other clinical trials \cite{ferreira2020use, beal2022comparing, berwanger2022sacubitril, romiti2023mobile, teramoto2024win} as an alternative to traditional TTFE approaches.

Given the increasing popularity of WR analysis, methods to estimate its statistical power have been developed \cite{Mao2021, yu_sample_2022, bonner2025power}. It was demonstrated for the HYPOECMO trial that WR analysis can outperform logistic regression in terms of power \cite{monzo2024use}. Additionally, simulations investigating the power of WR for composite endpoints involving TTE outcomes, alongside the Cox model, suggested that the WR generally had higher power. This was particularly evident when the highest-ranked outcome had a strong effect and high event rate, while the Cox model was more sensitive to the inclusion of additional components with small effects or high event rates \cite{wang2023statistical}.

Despite the growing interest in WR analysis, existing literature on its use in trial design remains limited in several aspects. Most studies focus on TTE outcomes and do not fully explore the impact of adding diverse outcome types into composite endpoints. Furthermore, there is limited guidance on sample size planning for precision-based trials using the WR. 

To address these gaps, we provide a comprehensive overview of WR analysis and its application at the design stage of clinical trials. We begin by introducing the WR and its key statistical properties. We then review existing approaches to power calculation and compare them to a simulation-based power analysis conducted for the planned IPHAK (Incidence of Primary Aldosteronism in patients with Hypokalemia) trial, also demonstrating how WR analysis can improve statistical power compared to single-endpoint analyses or compared to TTFE analysis. This result is further explored through Monte Carlo simulation studies. Finally, we discuss the use of the WR in precision-based trials and introduce a novel sample size formula based on the desired precision of the log-transformed WR, offering practical guidance for future trial planning.

\section{Methods}

\subsection{The Win Ratio}
The WR, introduced by Pocock et al. and derived from the Finkelstein–-Schoenfeld method, is designed to reflect the clinical hierarchy of outcomes within a composite endpoint \cite{Pocock2011}. It relies on hierarchical, pairwise comparisons between patients across the components of the composite, allowing outcomes of different types to be incorporated. In a two-arm trial comparing a treatment group ($T$) to a control group ($C$), patient pairs can be constructed using two main approaches:

\begin{itemize}
\item The \emph{risk-matched} approach forms pairs based on baseline risk profiles, yielding $\min\{N_T, N_C\}$ comparison pairs, where $N_T$ and $N_C$ are the sizes of group $T$ and $C$. Its validity depends on a predefined and unbiased matching procedure. To address imbalanced group sizes, adjustments to preserve power and precision were proposed \cite{Pocock2011}.

\item In practice, defining an unbiased matching procedure can be difficult, making the \emph{unmatched} approach more common \cite{monzo2024use}, especially in power calculations \cite{Mao2021, yu_sample_2022, bonner2025power}. This method compares all $N_T \times N_C$ possible pairs, avoiding matching complexity but risking bias if baseline risk distributions differ. While randomization typically mitigates this concern, it can be problematic in observational studies  \cite{monzo2024use}. Moreover, when strong prognostic variables exist, the unmatched approach typically yields lower power than the risk-matched approach \cite{Mao2021}.
\end{itemize}

\noindent Given pairs of patients from the treatment and control group, the following hierarchical evaluation is performed for each comparison pair:

\begin{enumerate}

\item Compare the two patients on the highest-ranked outcome in the hierarchy. If the patient in the treatment group has a more favorable outcome, the comparison is recorded as a win; if the outcomes are equal, it is a tie; otherwise, it is a loss.

\item If the comparison on the highest-ranked outcome results in a tie, proceed to compare the patients on the next outcome in the hierarchy. Continue this sequential evaluation until a win or loss is determined, or all outcomes have been considered.

\item If no win or loss is identified after comparing all outcomes, record the pairwise comparison as a tie.

\end{enumerate}

\begin{figure}
\centering

\includegraphics[width=0.9\textwidth]{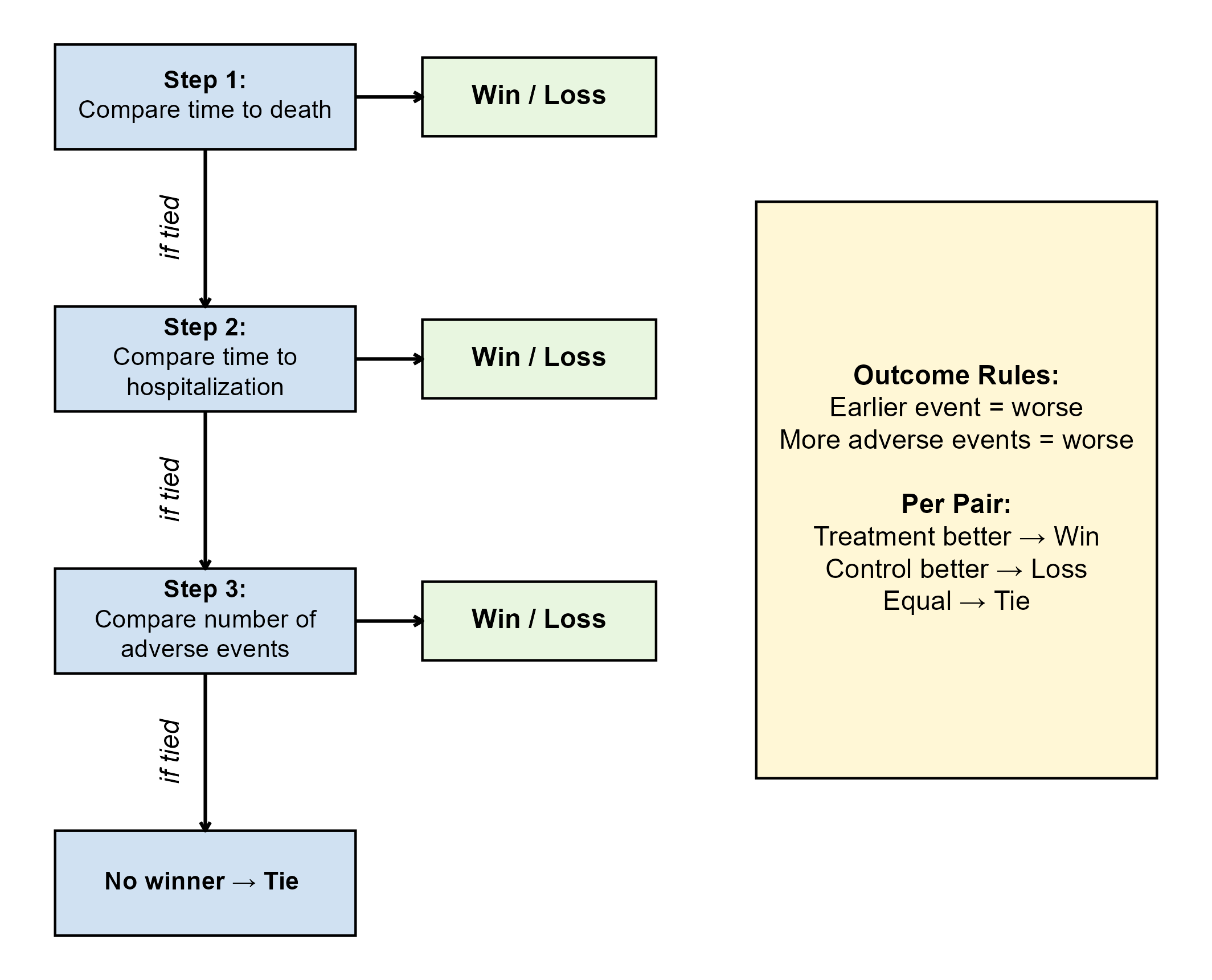}
\caption{Flowchart illustrating the hierarchical win ratio comparison for a single patient pair. The comparison begins with the most severe outcome --- time to death --- and proceeds sequentially through time to hospitalization and number of adverse events, if necessary. For each outcome, a win is record if the patient on treatment has a more favorable (i.e., later, absent, or less) event time or number of events. If the patient from the control group has a more favorable outcome, it is a loss. If no winner can be determined at a given level of hierarchy, the comparison continues to the next outcome in the hierarchy. If no difference is found across all outcomes, the pair is recorded as a tie.}
\label{fig:flow}
\end{figure}

\noindent Consider the following example: a two-arm parallel-group trial with a composite endpoint consisting of (1) time to death, (2) time to hospitalization, and (3) the total number of adverse events during follow-up. Unlike approaches that consider only the time to the first adverse event, this incorporates the total count of such incidents, providing a more comprehensive measure of patient experience. Each pair is compared hierarchically: first by time to death --- if one patient dies earlier or only one dies, the comparison is decided at this stage. If tied (e.g., neither dies during follow-up), the next outcome, hospitalization, is considered. If still unresolved, the number of adverse events is evaluated. Pairs not yielding a win or loss after all outcomes are recorded are ties. This stepwise decision process is illustrated in Figure~\ref{fig:flow}.

The WR is the ratio of the probability of a win to the probability of a loss, ranging from zero to infinity, with values above one favoring the treatment group:

\begin{equation*}
\WR = \frac{\text{P(Win)}}{\text{P(Loss)}} = \frac{N_{Win}}{N_{Loss}},
\end{equation*}

\noindent where $N_{Win}$ and $N_{Loss}$ denote the numbers of wins and losses. Under the null hypothesis of no treatment effect, the expected WR is one. Tied comparisons are excluded from the calculation, which can lead to information loss and, when ties are frequent, may result in overestimation of the effect size \cite{monzo2024use}. To address this, one can add a continuous endpoint at the end of the hierarchy, which, even if weakly associated with treatment, can help break ties and increase statistical power \cite{yu_sample_2022}. An alternative solution to solving ties is to use the win odds rather than the WR, in which the numerator and denominator are defined as the number of wins and losses increased by 50\% of ties, to include all patients pairs into the analysis \cite{brunner2021win}.

The WR can also be applied when outcomes are subject to censoring. In such cases, pairwise comparisons are only made when sufficient data are available for both patients. Specifically, for TTE outcomes, the comparison is restricted to the time interval during which neither patient is censored. Moreover, generalizations have been proposed, including stratified WR analysis \cite{dong2018stratified, dong2023stratified} and covariate adjusted WR analysis \cite{gasparyan2021adjusted, wang2025adjusted}.

The WR provides a nonparametric, model-free effect size that does not rely on assumptions like PH and remains valid regardless of the composite's structure \cite{Mao2021}. It can be interpreted as the odds that a randomly selected patient on treatment has a more favorable outcome than one from the control group, conditional on the comparison not resulting in a tie \cite{Pocock2011}. In simple terms, it reflects the relative likelihood of benefit under treatment \cite{Mao2021}. Conversion to the probability scale is straightforward if preferred.

However, the WR interpretation has limitations: it does not convey absolute effects --- for instance, a large WR could result even if the treatment delays a fatal event by on average only one week, which may lack clinical relevance \cite{monzo2024use}. Introducing efficacy margins into pairwise comparisons can address this but may increase the number of ties and reduce statistical power. A further limitation is that if the composite endpoint does not reflect clinical importance or real-world practice, the resulting effect estimates may be invalid. Sensitivity analyses are recommended when there is uncertainty about the composition and hierarchy of the composite \cite{monzo2024use}.

\subsection{Statistical Tests for the Win Ratio}

Pocock et al. propose the following statistical test for the risk-matched WR \cite{Pocock2011}; define the proportion of wins among informative comparisons (i.e., excluding ties) as:

\begin{equation*}
\phi_{{Win}} = \frac{N_{Win}}{N_{Win} + N_{Loss}}, \quad \text{with variance} \quad \Var(\phi_{Win}) = \frac{\phi_{Win} \cdot (1 - \phi_{Win})}{N_{Win} + N_{Loss}}.
\end{equation*}
\vspace{0.0cm}

\noindent Confidence intervals for $\phi_{Win}$ (i.e., $[\phi_{Win}^{\text{lower}}, \phi_{Win}^{\text{upper}}]$) may be obtained using methods such as the Wald or Wilson approaches, with corresponding WR confidence intervals derived via back-transformation. That is, since $\WR = h(\phi_{Win}) = \phi_{Win}/(1 - \phi_{Win})$, the limits of a WR confidence interval can be obtained as:
\[
\text{CI}_{\WR} = \left[ \frac{\phi_{Win}^{\text{lower}}}{1 - \phi_{Win}^{\text{lower}}},\; \frac{\phi_{Win}^{\text{upper}}}{1 - \phi_{Win}^{\text{upper}}} \right].
\]

\noindent Alternatively, we can obtain the approximate WR variance and standard error by applying the delta method:
\begin{align*}
\Var(\widehat{\WR}) &\approx \left( h'(\phi_{Win}) \right)^2 \cdot \Var(\phi_{Win}) = \frac{\phi_{Win}}{(1 - \phi_{Win})^3 (N_{Win} + N_{Loss})}, \\
\se(\widehat{\WR}) &\approx \sqrt{\Var(\widehat{\WR})} 
= \sqrt{\frac{\phi_{Win}}{(1 - \phi_{Win})^3 (N_{Win} + N_{Loss})}}.
\end{align*}

\noindent However, for hypothesis testing, the Wald test based on the log-transformed WR (\( \log(\widehat{\WR}) \)) is more efficient, as its sampling distribution more closely approximates a normal distribution \cite{yu_sample_2022}. The variance and standard error of \( \log(\widehat{\WR}) \) can be approximated by:
\begin{align*}
\Var(\log(\widehat{\WR})) &\approx \left( \frac{1}{\phi_{Win} (1 - \phi_{Win})} \right)^2 \cdot \Var(\phi_{Win}) 
= \frac{1}{\phi_{Win} (1 - \phi_{Win}) \cdot (N_W + N_L)}, \\
\se(\log(\widehat{\WR})) &\approx \frac{1}{\sqrt{\phi_{Win} (1 - \phi_{Win}) (N_W + N_L)}}.
\end{align*}

\noindent Based on this, the null hypothesis that the WR equals a specified value \( \WR_0 \) (\( H_0 : \WR = \WR_0 \)) can be tested using the statistic:
\[
Z = \frac{\log({\widehat{\WR}}) - \log(\WR_0)}{\se\left( \log({\widehat{\WR}}) \right)},
\]
\noindent which approximately follows a standard normal distribution under \( H_0 \), assuming a sufficiently large sample size so that the central limit theorem applies. 

In contrast, for the unmatched WR approach, the variance estimation is more complex due to dependency between pairwise comparisons. As a result, the use of bootstrap resampling procedures was proposed to obtain valid inference \cite{Pocock2011}.

\subsection{Power Calculation for the Win Ratio}
% However, real-world constraints such as logistical, financial, or ethical limitations often fix the sample size, shifting attention towards power estimation at a maximally affordable sample size.
In trial planning, investigators typically determine the sample size needed to achieve a desired power at a given significance level $\alpha$. Power calculation for clinical trials based on the risk-matched WR can be challenging, particularly when integrating relevant parameters into simulation designs. Most existing (closed-form) methods focus on the unmatched WR. To our knowledge, three methods have been proposed for estimating statistical power of the WR to this date: (1) one applies an approximate variance formula for $\log(\widehat{\WR})$, requiring input on anticipated tie proportion and group size ratio \cite{yu_sample_2022}; (2) another utilizes pilot data to estimate the joint outcome distribution \cite{Mao2021}; and (3) a third is a rank-based simulation method, applicable to composite endpoints with up to two outcomes and requiring iterative procedures for sample size determination \cite{bonner2025power}. In practice, however, we found the available implementation to be unreliable, consistently returning power estimates of either zero or one. A detailed description of these approaches is provided in the \textbf{Supplementary Material}.

Therefore, in designing a real-world clinical trial, we chose a simulation-based approach that allows specification of parameters in meaningful terms for clinical researchers. This simulation, informed by both statistical and clinical expertise, was carefully designed to estimate the power of the WR analysis compared to single-endpoint analyses. 

\section{Case Study: IPHAK Trial}\label{ch:iphak}

The IPHAK trial is a planned multicenter, randomized, controlled diagnostic study evaluating whether automated screening for Primary Aldosteronism (PA) in patients with hypokalemia improves diagnosis and clinical outcomes. The primary endpoint is a hierarchical composite outcome including: (1) binary classification of elevated blood pressure (EBP) based on WHO criteria; (2) change in defined daily dose (DDD) in blood pressure drugs.

The prevalence of PA in the study population was assumed to be 30\%. No treatment effect was expected among patients without PA (70\%), with any benefit confined to those with PA. Based on recruitment constraints, a total of 600 patients (300 per group) was considered feasible by clinical experts. After accounting for a 15\% drop-out rate, 255 patients per group are expected for analysis.

EBP and DDD were assumed to be uncorrelated, based on scenario analyses supporting a conservative approach. Among PA patients, 68\% in the treatment and 85\% in the control group were expected to have EBP at follow-up; for non-PA patients, the rate was assumed to be 81\% in both groups. The expected average change in DDD among treated PA patients is –1.364, reflecting a mix of surgical and medical management, while no change was expected in PA patients in the control group and in all non-PA patients. The correlation between baseline and follow-up DDD was assumed to be 0.5, with a standard deviation of 1.8. Lower values are considered favorable for both endpoints.

Based on these specifications, we conducted a simulation to estimate the power of the unmatched WR analysis, in comparison to standard analyses of either EBP (using a $\chi^2$-test) or DDD (using a $t$-test). EBP was simulated using a binomial distribution, while DDD was generated from a normal distribution. The simulation code is publicly available on \textbf{OSF (\href{https://osf.io/6qjup/}{https://osf.io/6qjup/})}.

\begin{figure}
\centering
\begin{knitrout}
\definecolor{shadecolor}{rgb}{0.969, 0.969, 0.969}\color{fgcolor}

{\centering \includegraphics[width=1\linewidth]{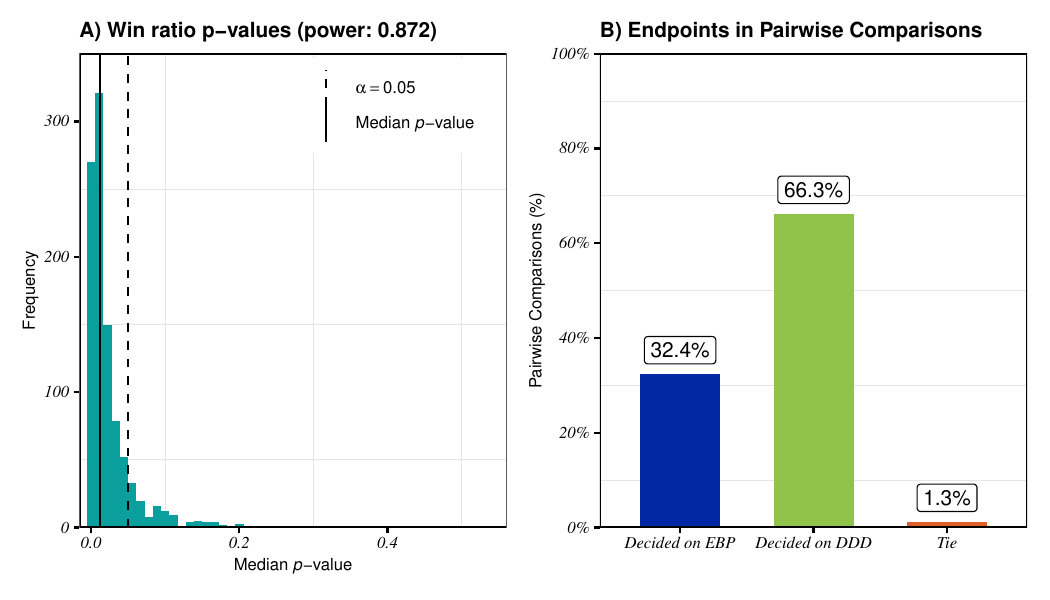} 

}

\end{knitrout}
\caption{Results from power calculation of the win ratio for the IPHAK trial: (A) Distribution of $p$-values across 1000 iterations; (B) Distribution of endpoints on which pairwise comparison were decided.}
\label{fig:poweriphak}
\end{figure}

The simulation results are shown in Figure~\ref{fig:poweriphak}. Approximately one-third of pairwise comparisons were resolved on EBP, while about two-thirds were decided on DDD. The estimated power of the unmatched WR analysis was 0.872, with a Monte Carlo standard error (MCSE) of 0.0106 based on 1000 simulation iterations. In contrast, analyzing only EBP or only DDD as the primary endpoint yielded powers of 0.298 and 0.726, respectively. These results suggest that the WR approach offers a substantial increase in power compared to analyses of the individual components of the composite endpoint.

From the simulation, we estimated an expected WR of 1.32. Using this estimate, along with the observed proportion of ties, the power estimation method using the approximate variance of the logarithmized WR \cite{yu_sample_2022} yields a power of 0.764, which is substantially lower than the empirical power observed in our simulation. Although the exact reason for this substantial difference is not fully clear, it is most likely due to the variance approximation of the log-transformed WR employed. The approach by Mao et al. could not be applied because pilot data was not available \cite{Mao2021}.

\section{Simulation}
\subsection{Win Ratio versus Single Outcome Analysis}

We observed a substantial increase in the expected power of the WR analysis for a composite endpoint combining a binary and a continuous outcome, compared to standard single-endpoint analyses. We investigate this further through a simulation study, structured according to the ADEMP framework \cite{morris2019using}.

\subsubsection{Aims}

We investigate whether WR analysis can increase statistical power when analyzing hierarchical composite outcomes, compared to analyzing only the highest-ranked individual outcome within the composite endpoint.

\subsubsection{Data-Generating Mechanism}

We consider a two-arm comparison between a new drug treatment and standard of care (SoC) with respect to a hierarchically ordered composite endpoint composed of a binary and a continuous outcome. The sample sizes in both groups are fixed at \( n_{T} = n_{SoC} = 20 \), reflecting a realistic scenario in which recruitment is limited. The effect size for the continuous outcome is varied using Cohen’s \( \delta \in \{0.1, 0.25, 0.5, 0.75, 1\} \), with higher values corresponding to more favorable outcomes. The proportion of favorable binary outcomes in the SoC group is fixed at \( p_{SoC} = 0.3 \), while the proportion in the treatment group varies over \( p_{T} \in \{0.35, 0.4, 0.5, 0.6, 0.7\} \). Continuous outcomes are generated from a normal distribution, and binary outcomes are drawn from a binomial distribution. In addition, we vary the hierarchical order of the outcomes, i.e., the sequence in which outcomes are compared in pairwise comparisons. All factor levels are varied in a full-factorial design.

\subsubsection{Estimands and Other Targets}

We investigate the estimated power across the presented scenarios, considered a target \cite{morris2019using}.

\subsubsection{Methods}

We compare the unmatched WR analysis for the composite endpoint to the $t$-test for the single analysis of the continuous outcome, and to Fisher's exact test for the single analysis of the binary outcome. We remark that the data-generating mechanism meets the distributional assumptions of both the $t$-test and Fisher's exact test, while the WR analysis makes no such assumptions.

\subsubsection{Performance Measures}

We investigate the estimated power of the different approaches under varying scenarios. The power in setting $s$ is:
\begin{equation*}
\text{Power}_s = \frac{1}{\niter} \sum_{i = 1}^{\niter} \mathbb{I}\{p_i \le \alpha\},
\end{equation*}

\noindent that is, we consider the proportion of $p$-values falling below a two-sided significance level $\alpha = 5\%$ across $\niter$ iterations. We report the estimated power together with its MCSE, which provides the basis for the estimation of the required number of iterations.

\subsubsection{Number of Iterations}

We aim to estimate the power sufficiently precise, requiring its MCSE to be $\le$ 1\%. The MCSE of the power in setting $s$, estimated from $\niter$ iterations, is:
\begin{equation*}\label{eq:sep}
\semc(\text{Power}_s) = \sqrt{\frac{\text{Power}_s \cdot (1 - \text{Power}_s)}{\niter}}.
\end{equation*}

\noindent This MCSE is maximized in case the power is 50\%, so using $\text{Power}_s = 0.5$, the required number of iterations to achieve a maximum MCSE $\le$ 1\% is:
\begin{equation*}
\niter =\frac{\text{Power}_s \cdot (1 - \text{Power}_s)}{\semc^2} = 2500.
\end{equation*}

\subsubsection{Results}

\begin{figure}[ht!]
\centering
\begin{knitrout}
\definecolor{shadecolor}{rgb}{0.969, 0.969, 0.969}\color{fgcolor}

{\centering \includegraphics[width=\maxwidth]{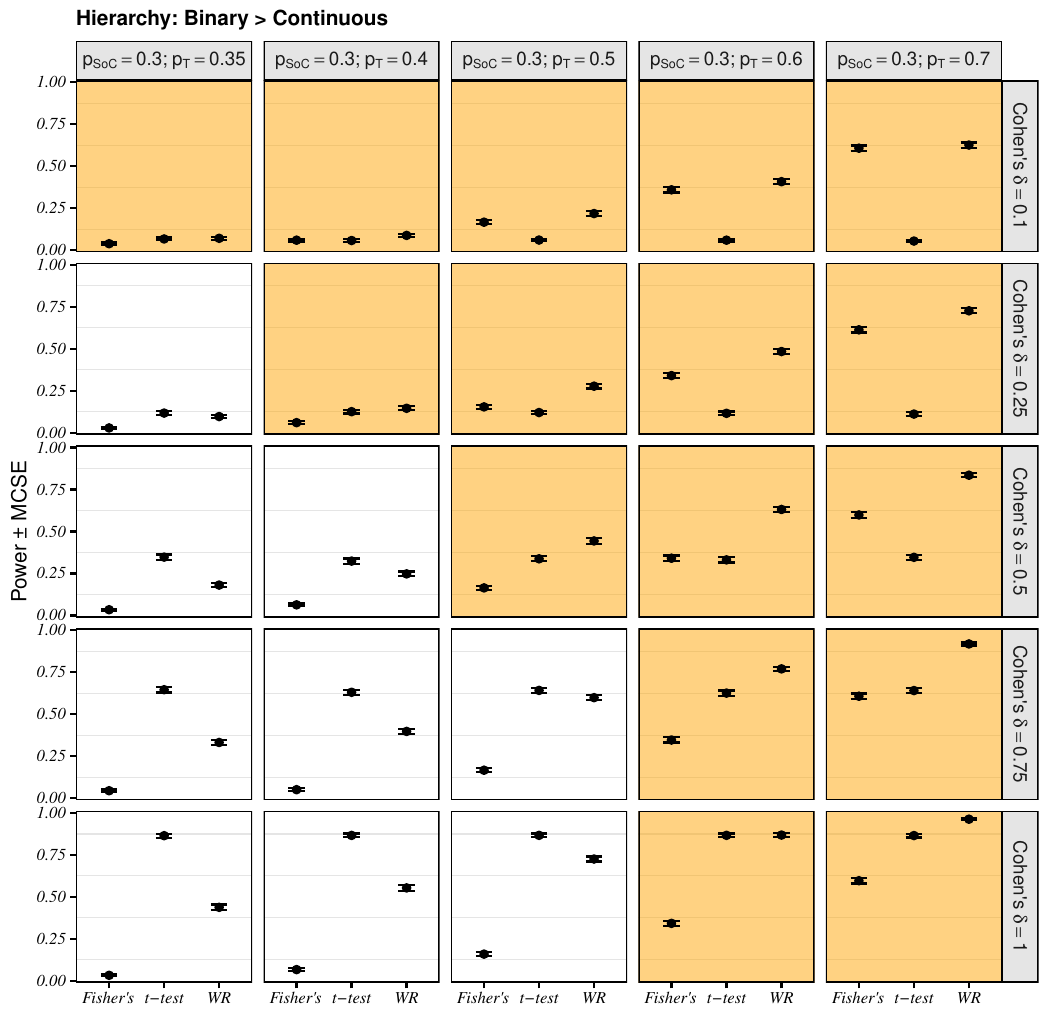} 

}

\end{knitrout}
\caption{Power comparison of the win ratio (WR) analysis for a composite endpoint, the $t$-test for the continuous outcome, and Fisher's exact test for the binary outcome, across varying effect sizes in both outcomes. The binary outcome is prioritized over the continuous outcome in the hierarchical analysis. Colored panels indicate scenarios in which the WR approach outperformed both other methods. Errobars represent Monte Carlo standard errors (MCSE).}
\label{fig:powerbc}
\end{figure}

Simulation code and results are available on \textbf{OSF (\href{https://osf.io/vzqe6/}{https://osf.io/vzqe6/})}. No non-convergences occurred. Figure~\ref{fig:powerbc} displays the estimated power when the binary outcome is ranked higher in the hierarchy than the continuous outcome. Colored panels correspond to settings where the WR approach achieved higher power than either of the other two approaches. We observe a clear pattern: the WR approach is advisable when the groups cannot be substantially better distinguished based on the lower-ranked continuous outcome as compared to the higher-ranked binary outcome. Interestingly, even when the risk difference is large and the difference in the continuous outcome is minimal, the WR approach still achieves greater power than Fisher's exact test. 

\begin{figure}[ht!]
\centering
\begin{knitrout}
\definecolor{shadecolor}{rgb}{0.969, 0.969, 0.969}\color{fgcolor}

{\centering \includegraphics[width=\maxwidth]{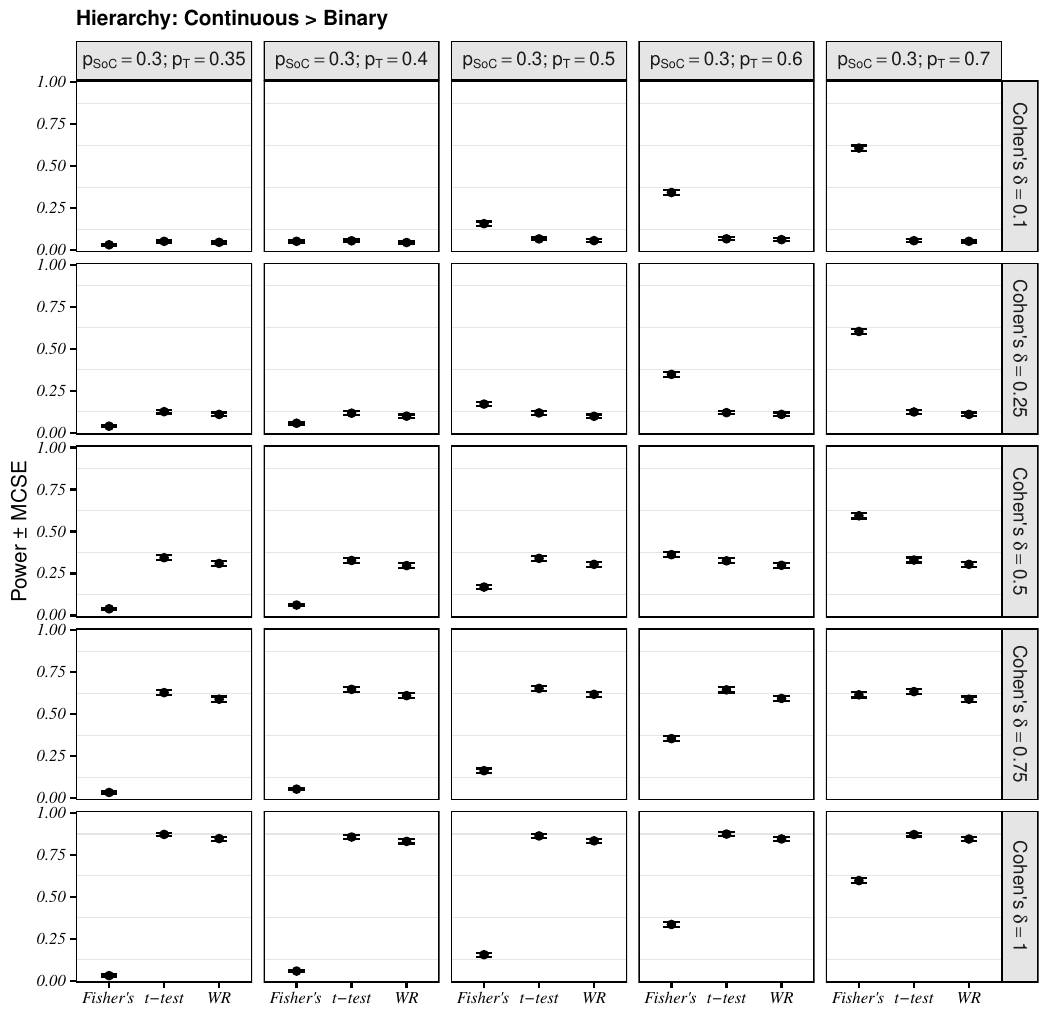} 

}

\end{knitrout}
\caption{Power comparison of the win ratio (WR) approach for a composite endpoint, the $t$-test for the continuous outcome, and Fisher's exact test for the binary outcome, across varying effect sizes in both outcomes. The continuous outcome is prioritized over the binary outcome in the hierarchical analysis. Colored panels indicate scenarios in which the WR approach outperformed both other methods. Errobars represent Monte Carlo standard errors (MCSE).}\label{fig:powercb}
\end{figure}

Figure~\ref{fig:powercb} shows the results when the continuous outcome is prioritized over the binary outcome. As expected, the WR approach never achieves higher power than both tests based on the individual outcomes. Even when the difference in the continuous outcome between groups is small, all decisions are still driven by it. Nevertheless, the power of the WR remains consistently only slightly lower than that of the $t$-test. In practice, this issue can be addressed by introducing efficacy margins on the continuous outcomes. This approach increases the number of ties on the continuous scale, thereby shifting more pairwise comparisons to be resolved based on the second-ranked binary outcome.

\subsection{Win Ratio versus Time-to-First-Event Analysis}

We conduct a simulation study to compare the statistical power of unmatched WR and time-to-first-event (TTFE) analyses. The simulation is conducted using a setup analogous to that for the composite endpoint consisting of one binary and one continuous outcome; therefore, we restrict the description to the key alterations made to accommodate the TTFE framework.

We consider the setup of a two-arm trial, comparing a new drug treatment with a SoC treatment. The composite endpoint consists of two hierarchically ordered TTE outcomes: (1) time to death and (2) time to hospitalization. We assume a follow-up period of two years (730 days), and an allocation ratio of 1:1 for a total sample size of 210 patients, resulting in 105 patients per group.

We use a Weibull model to generate event times, a standard approach \cite{bender2005generating}. We make the following assumptions: The expected survival rate in the SoC group is 70\% at the end of the follow-up period, corresponding to a two-year event rate of 30\%. Hence, denoting the survival time in the SoC group by the random variable $T_{\text{SoC}}$, we require that
\begin{equation*}
P(T_{\text{SoC}} \ge 730 \ \text{days}) = S_{\Wb}(730 \ \text{days}) = 0.7,
\end{equation*}

\noindent where $S_{\Wb}(\cdot)$ denotes the survival function of the Weibull model. The survival function of the Weibull model $\Wb(\lambda, \kappa)$ with shape parameter $\lambda \in (0, \infty)$ and scale parameter $\kappa \in (0, \infty)$ is:
\begin{equation*}
S_{\Wb}(t) = \exp\left(-\left(\frac{t}{\lambda}\right)^\kappa\right),
\end{equation*}

\noindent where here $t = 730$ days is the maximum follow-up time. Fixing $\kappa = \kappa'$, we obtain
\begin{equation*}
\lambda = \frac{t}{\left(-\log(S_{\Wb}(t))\right)^{\frac{1}{\kappa'}}}.
\end{equation*}
\vspace{0cm}

\begin{figure}[ht!]
\centering
\begin{knitrout}
\definecolor{shadecolor}{rgb}{0.969, 0.969, 0.969}\color{fgcolor}

{\centering \includegraphics[width=0.9\linewidth]{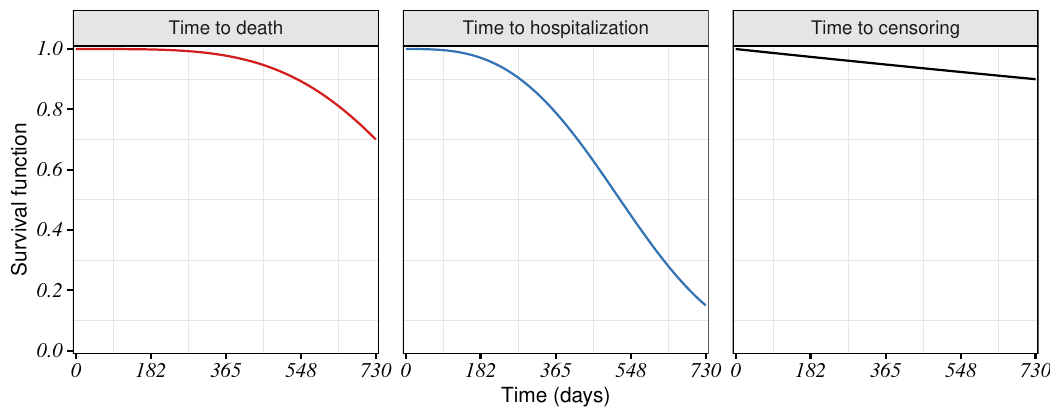} 

}

\end{knitrout}
\caption{Survival functions for the three data-generating distributions used to simulate event and right-censoring times in the standard of care group. Time to death is modeled using a Weibull distribution, calibrated to reflect a 70\% survival probability at 730 days. The second models time to hospitalization, also using a Weibull distribution, assuming that 85\% of patients are hospitalized at least once by 730 days. The third distribution models dropout times using an exponential distribution, corresponding to a 10\% overall dropout rate during the follow-up period.}
\label{fig:wbs}
\end{figure}

\noindent We choose $\kappa$ = 4 inducing $\lambda = $ 944.61. For time to hospitalization, we apply the same logic. We assume that at the end of the follow-up period, 85\% of patients in the SoC group were hospitalized at least once. Hence, a useful model is a Weibull distribution with $\kappa = 2$ and $\lambda =$ 530. Further, we consider a drop-out rate of 10\%, and model the time of right-censoring using an exponential distribution with $\lambda = $ 6928.59. The survival functions of these three data-generating distributions for the SoC group are displayed in Figure~\ref{fig:wbs}. For the analysis, we assume that event times are measured with a precision of one day.

\begin{figure}[ht!]
\centering
\begin{knitrout}
\definecolor{shadecolor}{rgb}{0.969, 0.969, 0.969}\color{fgcolor}

{\centering \includegraphics[width=\maxwidth]{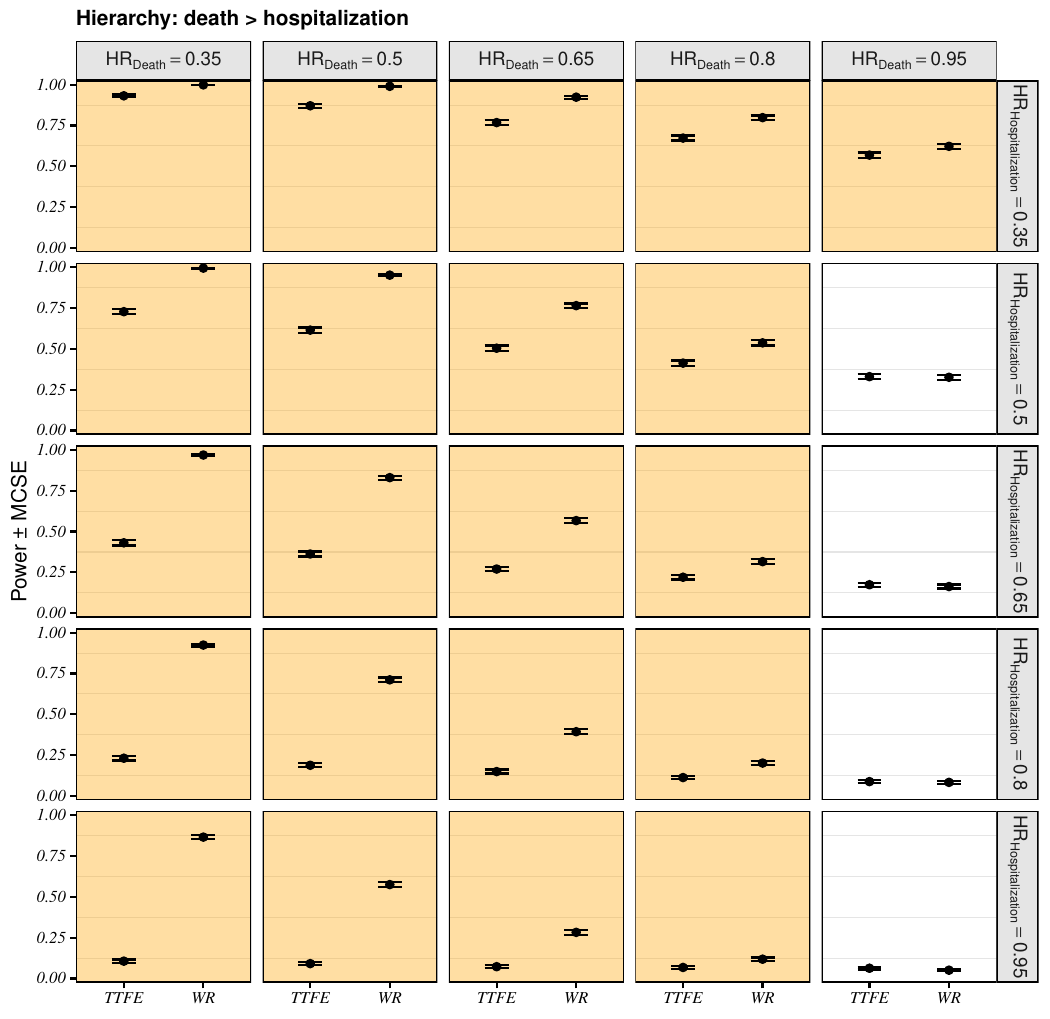} 

}

\end{knitrout}
\caption{Estimated power of the win ratio (WR) and time-to-first-event (TTFE) analyses across varying hazard ratios (HR) for both time-to-event outcomes in the composite endpoint. The power of the TTFE analysis is estimated using the log-rank test for the Cox proportional hazards model. Colored panels indicate scenarios in which the WR analysis outperformed the TTFE analysis.}
\label{fig:simrtte}
\end{figure}

We assess varying treatment effects of the new intervention compared to SoC, expressed as hazard ratios (HRs) within the Weibull model framework. Specifically, we examine HRs of 0.35, 0.5, 0.65, 0.8, and 0.95 for both outcomes comprising the composite endpoint. Censoring probabilities are assumed to be equal across treatment groups. For the TTFE analysis, we use the Cox PH model with the score-based log-rank test. We vary the effect sizes in a full-factorial design and perform 2500 simulation iterations to obtain sufficiently precise power estimates. We display the results in Figure~\ref{fig:simrtte}.

We generally observe that the WR analysis has greatly increased power to TTFE analysis under scenarios where the effect on death, that is, the first outcome in the hierarchy, is not minimal. We observe increases in power in the WR analysis compared to the TTFE analysis of over 50\%. Under scenarios where the HR quantifying the treatment effect on time to death only marginally differs from one (corresponding to almost no effect), the WR approach performs as poorly with respect to power as TTFE analysis. Generally, under no scenarios WR analysis yielded a power which was substantially smaller than the power of TTFE analysis.

\section{Sample Size Planning for the Win Ratio in Precision-Based Trials}

Power-based trials are often designed primarily to achieve statistical significance --- that is, a $p$-value below a conventional threshold, typically 5\%, for the primary outcome \cite{bland2009tyranny}. If this threshold is met, the null hypothesis is rejected; otherwise, it is retained. This binary accept-or-reject framework has been widely criticized, as many studies are not directly tied to decision-making and do not require strict classification into "significant" or "non-significant" results \cite{stang2010ongoing, Fraser2019}. Practical constraints, such as limited resources or recruitment challenges, may also make adequate powering unrealistic, potentially setting studies up for failure. Even when power is sufficient, this approach can mislead: clinically trivial effects may be deemed "significant," while meaningful effects may be dismissed as "non-significant" due to overpowered or, conversely, underpowered analyses \cite{van2017statistical}. To address these issues, the focus should shift from achieving statistical significance to evaluating the clinical relevance of effects.

An alternative is given by precision-based trials, which aim to estimate effects with a predefined level of precision rather than assessing statistical significance \cite{Julious2023}. In the design stage, sample size calculations are performed to ensure that such precision, typically captured by the width of the confidence interval of the estimator, is achieved. In contrast to power-based trials, where the sample size is chosen to detect an effect of a certain size with a certain power (e.g., 80\%), the actual effect size plays a secondary role in the sample size planning of precision-based trials.

With respect to the unmatched WR analysis, we employ the approximate variance of $\log(\widehat{\WR})$ \cite{yu_sample_2022} to construct a formula for the total sample size based on the width of a Wald confidence interval. A two-sided $(1-\alpha)\cdot 100$\% Wald confidence interval for $\log(\WR)$ can be constructed as:
\begin{equation*}
\log(\widehat{\WR}) \pm Z_{1-\alpha/2} \cdot \se(\log(\widehat{\WR})),
\end{equation*}

\noindent where $Z_q$ denotes the $q$th quantile of a standard normal distribution. A confidence interval for the WR can be obtained by applying the exponential transformation. Hence, for a specified total width of the Wald confidence interval for $\log(\WR)$ --- potentially obtained from the desired width of the WR confidence interval --- we can compute the required total sample size independent of the actually expected value of the WR. That is, we can re-equate the expression of the total width:

\begin{equation*}
\text{Width} = 2 \cdot Z_{1-\alpha/2} \cdot \sqrt{\frac{1}{N_{total}} \cdot  \frac{4 \cdot (1 + \ptie)}{3 \cdot \pt \cdot (1 - \pt) \cdot (1 - \ptie)}},
\end{equation*}

\noindent where $\pt$ denotes the probability of being in the treatment group, and $\ptie$ denotes the probability of a tie, for the total sample size to yield:
\begin{equation*}
N_{total} = \frac{16 \cdot Z_{1-\alpha/2}^2 \cdot (1 + \ptie)}{3 \cdot \pt \cdot (1 - \pt) \cdot (1 - \ptie) \cdot \text{Width}^2}.
\end{equation*}
\vspace{0cm}

\noindent Suppose our goal is to estimate $\log(\WR)$ with a total confidence interval width of 0.8. Assuming balanced group sizes ($\pt = 0.5$) and a low tie probability ($\ptie = 0.02$), this would require 67 patients per group.

While we acknowledge the limitations due to the approximation-based formula of the variance and the difficulty of choosing the expected proportion of ties (outlined in the \textbf{Supplementary Material}), precision-based trials are usually applied in exploratory stages of scientific research and generally subject to larger uncertainty with respect to effects. Generally, a slight decrease in precision is not as problematic as a slight decrease in power. The latter may imply that a trial is labeled "not significant" or "not successful", while such a strict dichotomy is typically not applied to precision-based trials.

\section{Discussion}

We investigated the WR for analyzing composite endpoints in clinical and pharmaceutical trials, focusing on power and sample size estimation for the design of both power-based and precision-based trials. Through a case study and corresponding simulation, we demonstrated that applying the WR analysis to a composite endpoint can result in greater statistical power compared to conventional single-endpoint methods. 

We examined this result through a simulation study, showing that the WR tends to outperform single-endpoint analyses when treatment effects on lower-ranked outcomes are moderate compared to those on higher-ranked outcomes --- provided the higher-ranked outcomes are not continuous, in which case all decisions are made high in hierarchy. Additionally, we conducted a simulation study to investigate the power of the WR analysis as compared to TTFE analysis. We found that the WR approach tends to greatly outperform TTFE analysis, provided the effect on the outcome highest in hierarchy is not minimal. This finding is consistent with that of earlier studies \cite{wang2023statistical}.

Finally, we discussed the potential of using the WR in precision based trials, and provided a simple formula to estimate the required sample size for a specified width of the confidence interval for the logarithmized WR. 

In terms of power and sample size estimations, previously introduced calculations \cite{Mao2021, yu_sample_2022} offer useful alternatives when simulation-based approaches are impractical. However, their applicability may be limited by the lack of pilot data and the difficulty of accurately specifying input parameters, as empirical estimates for the WR are still relatively scarce. In our case study, the analytic formula based on the approximate variance of the logarithmized WR \cite{yu_sample_2022} further underestimated the WR’s power compared to simulation results.

Our study has several limitations. First, we explored only a limited set of scenarios relevant to clinical research. The literature on simulating TTE data is extensive \cite{crowther2013simulating}, and many additional scenarios warrant investigation. For instance, we simulated PH data consistent with Cox model assumptions. However, the Cox model is frequently applied even when PH assumptions are violated \cite{dormuth2023comparative}; in such cases, the WR remains applicable and may offer increased power. 

Regarding outcome types, it would be valuable to extend our work to ordinal outcomes (e.g., as the mRS scale --- a relevant outcome in stroke research \cite{ospel2024pragmatic}), especially since both WR and common ordinal methods are non-parametric, whereas parametric approaches (e.g., proportional odds models) rely on strong assumptions.

Finally, we used fixed sample sizes to facilitate interpretation. In practice, however, power analyses must be tailored to study-specific assumptions, and researchers should conduct their own power calculations at design stage to reflect the specific assumptions of their study.

The presented ideas and simulations can be extended to stratified \cite{dong2018stratified, dong2023stratified} and covariate-adjusted WR analysis \cite{gasparyan2021adjusted, wang2025adjusted}. The ICH E9 guidelines support prespecified covariate adjustment to enhance precision and interpretability in randomized trials \cite{ICHE9}. Our simulations could incorporate predefined associations between covariates and outcomes to support such extensions.

Our findings highlight the potential of the WR to improve statistical efficiency in clinical trial design using composite endpoints, particularly when no single component dominates the treatment effect. However, when continuous outcomes occupy the top of the hierarchy, these tend to drive overall analysis, sidelining contributions of lower-ranked outcomes and limiting benefits of hierarchical win ratio analysis.

% ======================= Ethics and Integrity Policies ========================

\subsection*{Author Contributions}
DK: Conceptualization, methodology, software, validation, formal analysis, visualization, investigation, writing - original draft; MS: Investigation, writing - review \& editing; FB: Investigation, writing - review \& editing; UH: Conceptualization, methodology, validation, supervision, writing - review \& editing.

\subsection*{Data Availability Statement}
Simulation code and results for the presented case study are available on \textbf{OSF: \href{https://osf.io/6qjup/}{https://osf.io/6qjup/}}. Code and results for the simulation studies are also available: \textbf{\href{https://osf.io/vzqe6/}{https://osf.io/vzqe6/}}. No individual patient data were used in the preparation of this manuscript.

\subsection*{Funding Statement}
No funding was received for this article.

\subsection*{Conflict of Interest Disclosure}
The authors declare that they have no conflict of interest related to this article.

\subsection*{Ethics Approval Statement}
This study did not involve human participants or patient data and therefore did not require ethics approval.

\subsection*{Patient Consent Statement}
This study did not involve human participants or patient data and therefore did not require patient consent.

\subsection*{Permission to Reproduce Material from Other Sources}
No material from other sources was used or reproduced in this study.

% ============================== Appendix ======================================
\section*{Supporting information}

The following supporting information is available as part of the online article:

\noindent \textit{Summary of Proposed Power Calculations for Win Ratio Analysis.}

\appendix
\section{Summary of Proposed Power Calculations for Win Ratio Analysis}\label{app:power}
\subsection{Power Calculation for the Win Ratio by Yu et al. \cite{yu_sample_2022}}\label{sec:yu}

Calculating the variance of the unmatched win ratio (WR) is challenging due to the dependence among pairwise comparisons, necessitating resampling methods. Consequently, such calculations require access to patient-level data and are not well suited for standard power or sample size calculations. To address this limitation, the authors proposed a formula for the approximate variance of the estimated logarithmized WR ($\log(\widehat{\WR})$), which does not require individual patient data. Instead, its computation requires the proportion of patients assigned to the treatment group ($\pt$) and the anticipated probability of a tie between treatment and control pairs ($\ptie$):
\begin{equation*}
\Var(\log(\WR)) \approx \frac{1}{N_{total}} \cdot \left( \frac{4 \cdot (1 + \ptie)}{3 \cdot \pt \cdot (1 - \pt) \cdot (1 - \ptie)}          \right) = \frac{1}{N_{Total}} \cdot \sigma^2,
\end{equation*}
\vspace{0cm}

\noindent where $N_{total}$ denotes the total sample size. The approximation may be inaccurate when the true WR is either very large or small and tends to underestimate in small samples. Despite these limitations, assuming that $\log(\widehat{\WR})$ approximately follows a normal distribution with mean $\log(\WR)$ and above variance, standard sample size formulas can be applied. The required total sample size for a two-sided test can be estimated as:
\begin{equation*}
N_{Total} \approx \frac{\sigma^2 \cdot (Z_{1-\alpha/2} + Z_{1-\beta})^2}{\log(\WR)^2},
\end{equation*}
\vspace{0cm}

\noindent where $\alpha$ and $\beta$ refer to the type I and type II error rates, $Z_q$ denotes the $q$th quantile from the standard normal distribution and $\sigma^2$ needs to be estimated from the allocation ratio and the anticipated proportion of ties. The equation can be restructured to obtain the power:
\begin{equation*}
1 - \beta = 1 - \Phi \left(Z_{1-\alpha/2} - \log(\WR) \cdot \frac{\sqrt{N}}{\sigma}           \right),
\end{equation*}
\vspace{0cm}

\noindent where $\Phi(\cdot)$ denotes the cumulative distribution function (CDF) of the standard normal distribution. For a one-sided hypothesis test, the above formulas can be modified by substituting $\alpha$ in place of $\alpha/2$. Furthermore, the authors also derived an approximate variance formula for the stratified WR, which similarly enables sample size and power calculations in stratified WR analyses.

\begin{figure}[ht!]
\centering
\begin{knitrout}
\definecolor{shadecolor}{rgb}{0.969, 0.969, 0.969}\color{fgcolor}

{\centering \includegraphics[width=0.9\linewidth]{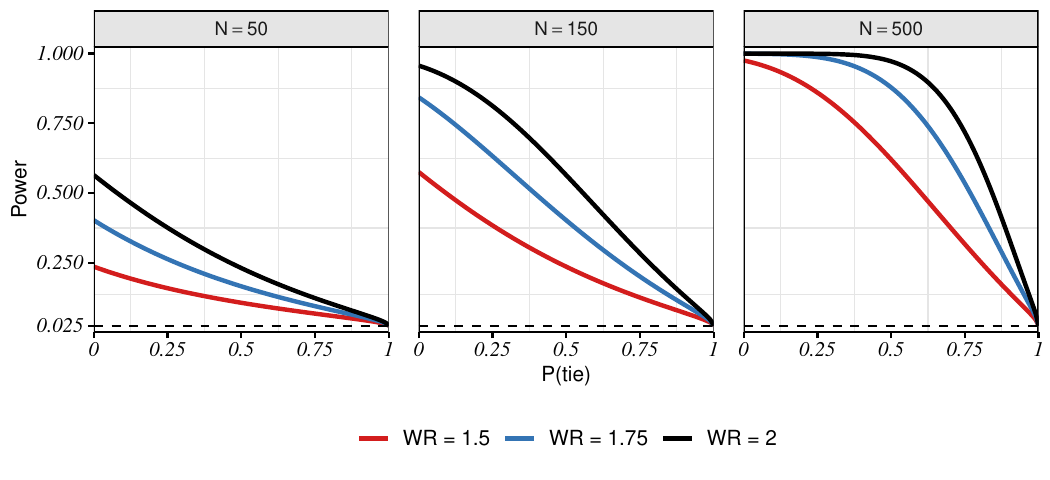} 

}

\end{knitrout}
\caption{Power calculation by \cite{yu_sample_2022} as a function of the proportion of ties for varying sample sizes (N) and win ratios (WR). The calculations assume a two-arm trial with a 1:1 allocation ratio and a two-sided significance level of $\alpha = 0.05$.}
\label{fig:yu2022ties}
\end{figure}

In practice, accurately estimating statistical power can be challenging. As shown, the power of the test is sensitive to the proportion of ties. Consider the following illustrative scenario: a two-sided test conducted at a significance level of $\alpha = 0.05$, with total sample sizes of 50, 150, or 500 patients, WRs of 1.5, 1.75, or 2, and a balanced 1:1 treatment allocation. Figure~\ref{fig:yu2022ties} shows how variations in the proportion of ties can markedly influence the statistical power of the test. This sensitivity is especially pronounced in smaller sample sizes, where misspecification of the tie proportion can lead to substantial under- or overestimation of power.

However, this issue virtually diminishes when the hierarchical composite endpoint includes at least one continuous outcome measured with adequate precision, as such outcomes typically greatly reduce the frequency of ties, also illustrated in the case study presented in the main text. This is consistent with earlier findings \cite{yu_sample_2022}. The authors further remark that introducing an additional continuous endpoint to resolve ties can enhance power --- even though it may reduce the estimated WR.

\subsection{Power Calculation for the Win Ratio by Mao et al. \cite{Mao2021}}

Similar to the previously introduced method, this method relies on the logarithmized WR and addresses estimation uncertainty through a noise parameter derived from rank-based statistics. It requires to specify the anticipated outcome distribution in the reference (control) group and uses the expected WR to derive the corresponding outcome distribution in the treatment group. This approach parallels standard sample size calculations for, e.g., binary outcomes, where event rates for both groups are specified. In practice, the event rate in the control group and a treatment effect measure (such as risk difference) are defined, from which the treatment group’s distribution is inferred.

Their approach can be summarized as follows: Let $W(y) = \text{P}(Y_C > y)$ and $L(y) = \text{P}(Y_C < y)$, where $Y_C$ denotes the outcome in the reference (control) group. Then, the function $R(y) := W(y) - L(y) \in [-1,1]$ defines a population-level generalized rank function. The variability of this function under the null hypothesis is captured by the standard rank variance:
\begin{equation*}
\xi_0^2 = \int R(y)^2 f_R(y) \ d y,
\end{equation*}

\noindent where $f_R(y)$ denotes the density function of the generalized rank function. Then, the total required sample size for a two-sided test can be approximated as:
\begin{equation*}
N_{total} = \frac{\xi_0^2 \cdot (Z_{1-\beta} + Z_{1-\alpha/2})^2}{\pc \cdot (1 - \pc) \cdot W_0^2\cdot \log(\WR)^2}
\end{equation*}
\vspace{0cm}

\noindent where $\pc$ denotes the proportion of patients in the reference group, $Z_q$ denotes the $q$th quantile of the standard normal distribution, $\alpha$ and $\beta$ represent the type I and type II error rates, respectively, and $W_0$ is the proportion of wins under the null hypothesis. Inverting this expression yields an estimate for the statistical power:

\begin{equation*}
1 - \beta = \Phi\left(\frac{W_0 \cdot \log(\WR) \cdot \sqrt{\pc \cdot (1-\pc)\cdot N_{total}}}{\xi_0} - Z_{1-\alpha/2}\right),
\end{equation*}
\vspace{0cm}

\noindent where $\Phi(\cdot)$ is the standard normal CDF. \cite{Mao2021} also propose expressions for estimating both $R(y)$ and $W_0$, provided the distribution $f_R(y)$ is specified.

The method by \cite{Mao2021} provides a more precise alternative to the approximation introduced by \cite{yu_sample_2022}. However, it depends critically on assumptions regarding the distribution of the generalized rank function under the null hypothesis, which may be difficult to specify in practice, especially in the absence of pilot data. Although statistically rigorous, the method is technically complex and may be challenging to implement without substantial prior knowledge.

\subsection{Rank-Based Simulation Approach for Power Calculation \cite{bonner2025power}}

The main limitations of the discussed approaches \cite{yu_sample_2022, Mao2021} are the reliance on the anticipated WR and proportion of ties or the dependence on availability of pilot data. To overcome these challenges, Bonner et al. propose a rank-based simulation method for power estimation of WR analyses. Their approach avoids assumptions about outcome distributions, relying instead on rank distributions, and is tailored for unmatched WR designs.

Their rank-based simulation approach proceeds as follows: At a given level of a composite endpoint, outcome comparison corresponds to comparison of ranks assigned based on the outcome values. Let \( X \) be a hypergeometric random variable representing the number of patients under treatment with ranks in the top half of the rank distribution (i.e., better outcomes). For equal sample sizes, under the null hypothesis of no treatment effect, $X$ has mean $\text{P(Win)} = \phi_{Win} = 0.5$. Under the alternative hypothesis, \( X \) follows a noncentral hypergeometric distribution with mean \( \phi_{Win} \). Hence, the key parameters to specify are the expected proportion of treatment wins at each stage of the composite endpoint. While this may seem more intuitive than specifying the anticipated WR, the two are equivalent --- since the proportion of treatment wins is directly related to the WR, subject to the proportion of ties.

However, \cite{bonner2025power} argue that the WR and the proportion of ties are additionally influenced by censoring, the distribution of ties for each component of the composite outcome, and the correlation between outcomes. These complexities make it difficult to anticipate the true WR and tie proportion in advance. Their approach, therefore, enables a more straightforward specification by focusing directly on the distribution of ranks. One may argue, however, that these same factors can also affect the win proportion in the treatment group, potentially obscuring the specification required for the simulation approach.

Once \( X = x \) is drawn, \( x \) ranks are sampled uniformly from the top half, i.e., from the discrete uniform distribution on \([1, N/2]\), and \( N - x \) ranks from the bottom half, i.e., from \([N/2 + 1, N]\). The remaining ranks are assigned to the control group. Ranks are then used to compute the WR, and a two-sided bootstrap confidence interval (default: 500 boostrap samples) is constructed to estimate the statistical power. The process is repeated for a number of iterations (default: 1000) to obtain a precise estimate of the power.

This approach can be extended to handle unequal group sizes, censoring, missing data, ties, and correlated ranks. Censoring or missingness is addressed by assigning the next available rank, though this may introduce bias. Rank correlations can be modeled using a Gaussian copula.

Evaluated in a simulation, the rank-based simulation approach yielded power results similar to those of the method based on the approximate variance of the logarithmized WR \cite{yu_sample_2022}, while however being computationally more intensive. The method is implemented in the \textsf{winratiopss} package (\href{https://github.com/LaurenBonner/winratiopss}{https://github.com/LaurenBonner/winratiopss}). The implementation allows for power estimation based on a composite endpoint consisting of maximum two endpoints, and does not directly allow for the estimation of the sample size.

% =============================== References ===================================
\newpage
\nocite{R}

\bibliography{../references.bib}

\vfill

\end{document}